# Strain Induced Indirect-to-Direct Bandgap Transition, Photoluminescence Enhancement, and Linewidth Reduction in Bilayer MoTe$_2$

*Yueyang Yu[1], Chuan-Ding Dong[2], Rolf Binder [3,&], Stefan Schumacher[2,3,\*], and Cun-Zheng Ning [1,4,$]*

[1]*School of Electrical, Energy, and Computer Engineering, Arizona State University, Tempe, AZ 85287, USA*

[2]*Department of Physics and Center for Optoelectronics and Photonics Paderborn*

*(CeOPP), Paderborn University, Paderborn 33098, Germany*

[3]*Wyant College of Optical Sciences, University of Arizona, Tucson, AZ 85721, USA*

[4]*New permanent address: Department of Electronic Engineering and Tsinghua International Center for Nano-Optoelectronics, Tsinghua University, Beijing 100084, China, cning@tsinghua.edu.cn; experiments performed at Arizona State University*

*Communication authors:*

[$] *cning@asu.edu,* [&] *binder@optics.arizona.edu,* [\*] *stefan.schumacher@uni-paderborn.de*

**Abstract**
Two-dimensional (2D) layered materials provide an ideal platform for engineering electronic and optical properties through strain control because of their extremely high mechanical elasticity and sensitive dependence of material properties on mechanical strain. In this paper, a combined experimental and theoretical effort is made to investigate the effects of mechanical strain on various spectral features of bilayer MoTe$_2$ photoluminescence (PL). We found that bilayer MoTe$_2$ can be converted from an indirect- to direct-bandgap material through strain engineering, resulting in a photoluminescence enhancement by a factor of 2.24. Over 90% of the PL comes from photons emitted by the direct excitons at the maximum strain applied. Importantly, we show that strain effects lead to a reduction of the overall linewidth of PL by as much as 36.6%. We attribute the dramatic decrease of linewidth to a strain-induced complex interplay among various excitonic varieties such as direct bright excitons, trions, and indirect excitons. Our experimental results on direct and indirect exciton emission features are explained by theoretical exciton energies that are based on first-principle electronic band structure calculations. The consistent theory-experimental trend shows that the enhancement of PL and the reduction of linewidth are the consequences of the increasing direct exciton contribution with the increase of strain. Our results demonstrate that strain engineering can lead to a PL quality of the bilayer MoTe$_2$ comparable to that of the monolayer counterpart. The additional benefit of a longer emission wavelength makes the bilayer MoTe$_2$ more suitable for Silicon-photonics integration due to the reduced Silicon absorption.

Layered transition metal dichalcogenides (TMDCs) have been widely studied for their potential application in photonic devices.[1],[2] One of the advantages of TMDCs has been their layer-number dependent bandgaps which lead to optical emission covering a large spectral range.[3],[4] However, most of TMDCs are indirect bandgap materials for the thickness of more than a monolayer.[3] Such indirect bandgap multilayer TMDCs have a low quantum yield and are poorly suited for applications in optoelectronic devices. Among all TMDCs, $MoTe_2$ has been especially interesting as one of the very few monolayer materials that emit in the near-infrared with photon energy slightly below the Silicon absorption edge, suitable for silicon-based integrated photonics. For many applications, emission of longer wavelengths than that of monolayer $MoTe_2$ would be desirable both to cover a different wavelength band and to be further away from the Silicon absorption edge to reduce the absorption. In this sense, a bilayer $MoTe_2$ with a direct bandgap would be ideal with less absorption.

Strain engineering has been applied on various layered TMDCs to modify their optical properties experimentally.[5]-[13] Since many monolayer TMDCs are direct-gap semiconductors, strain engineering typically degrades their PL qualities, leading to decreased intensities and increased linewidths due to the direct-to-indirect bandgap transition.[5]-[12] Among all TMDCs, strain-induced indirect-to-direct bandgap transition has only been reported on bilayer $WSe_2$ so far.[13] For bilayer $WSe_2$, the difference between the indirect and direct gap is around 40meV, making the crossover possible with achievable strain values. Bilayer $MoTe_2$ shares a similarly small energy difference between the indirect and direct gaps. It is estimated to be around 60meV by a theoretical calculation,[14] much smaller than in other bilayer TMDCs.[15]-[17] It was also theoretically predicted that bilayer $MoTe_2$ can be converted into a direct bandgap semiconductor under an isotropic tensile strain of 1% in three dimensions.[18] All this indicates that bilayer $MoTe_2$ is an ideal candidate for potentially changing from indirect to direct bandgap through strain engineering, leading to further increase of PL yield. In this work, we experimentally demonstrate simultaneous PL enhancement and linewidth reduction on bilayer $MoTe_2$ by applying uniaxial tensile strain, showing strong signatures of transition from an indirect to a direct bandgap material. Through theoretical calculations specifically incorporating the uniaxial strain features of our experiments, we were able to discover the interplay between different exciton species during the process of transition from indirect to direct bandgap.

**RESULTS AND DISCUSSION**

Experimental Results:

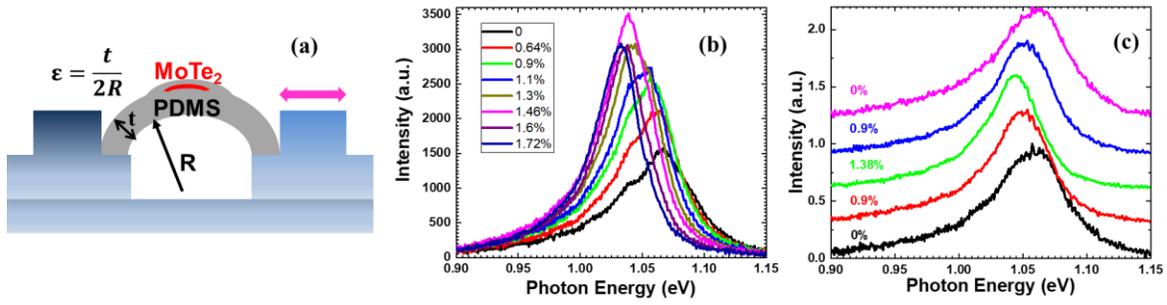

**Figure 1.** (a) Schematic of the set-up used for applying strain to MoTe$_2$. (b) Photoluminescence spectra of bilayer MoTe$_2$ under different levels of tensile strain. (c) Normalized PL spectra (offset for clarity) for successive levels of strain from 0%, 0.9%, 1.38%, 0.9%, and 0% to test the reversibility of the strain effects.

The customized set-up used to apply strain on MoTe$_2$ is schematically shown in Figure 1(a). The polydimethylsiloxane (PDMS) substrate is bent by fixing one end and pressing the other end by a moving block. Uniaxial tensile strain is transferred onto the sample from the PDMS. Its value is estimated as $\varepsilon = t/2R$, where t is the thickness of the PDMS and R is the radius of curvature. For the preparation of PDMS and the discussion on strain transfer, see Methods and Supporting Information S9, respectively.

In-situ PL measurement is performed as strain is applied on the samples. Figure 1b shows the PL spectra of bilayer MoTe$_2$ at different strain levels up to 1.72%. There are several direct observations as strain is increased. The PL peak red-shifts along with the increase of PL intensity as strain is increased. The maximum redshift is around 34.8meV in the range of strain applied with the PL peak intensity increasing by a factor of ~2.24. Additionally, we observe a reduction in the spectral linewidth. These features will be more systematically studied later on (see Figure 5d), together with the discussions of the physical origins of these changes. We also tested the reversibility of the strain effects by applying and releasing the strain. The normalized PL spectra during this process are shown in Figure 1c, where we can see a fully restored spectrum after strain is released. Based on the steady improvement of the PL quality with strain and the complete reversibility of spectral features, the material seems to be free from mechanical damage or slippage with respect to the substrate after up to 1.38% strain is applied, indicating full reversibility of the strain effects within this range.

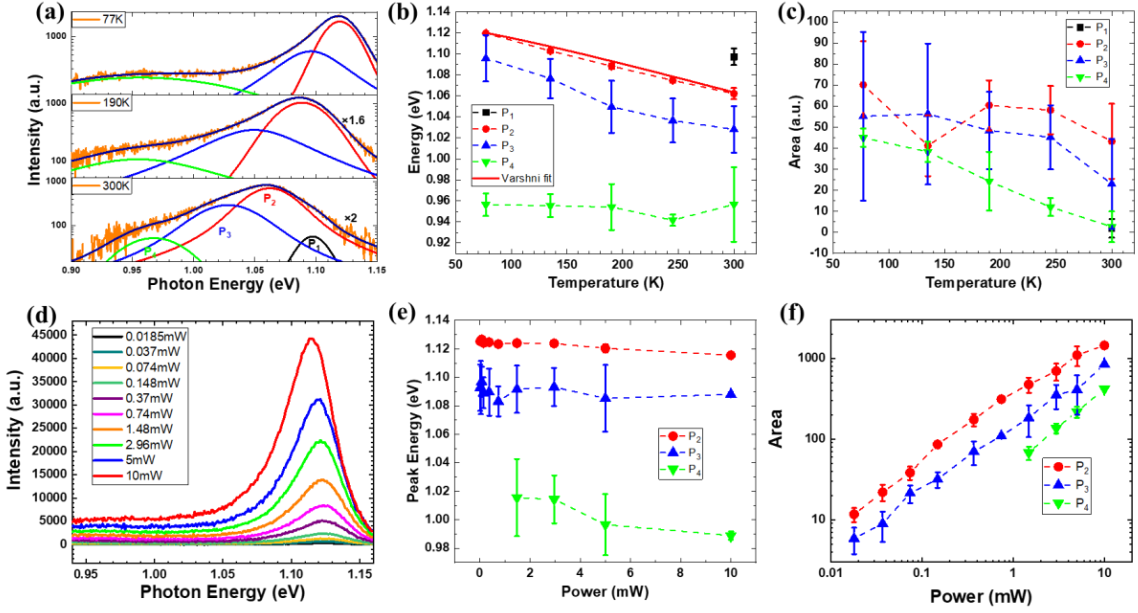

**Figure 2.** (a) PL spectra of bilayer MoTe$_2$ without strain at 77K, 190K, and 300K fitted with multiple Voigt peaks. (b) Peak energies vs. temperature, the energies of Peak 2 are fitted by the Varshni Equation. (c) Integrated peak intensities of Peak 1-4 vs. temperature. (d) PL spectra of unstrained bilayer MoTe$_2$ at different pumping levels at 77K. (e) Peak energies of Peak 2-4 vs. pumping power extracted from (d). (f) The light-out-light-in (L-L) curve: Integrated peak intensity of Peak 2-4 extracted from (d). The error bars in (b), (c), (e), (f) represent one standard deviation in the least-squares fits, details of which can be found in the Supporting Information S1.

To identify the spectral features of PL of bilayer MoTe$_2$, temperature-dependent PL measurements without applying strain was conducted. PL spectra were recorded at different temperatures down to 77K, shown in Figure 2a in log scale. The spectra are fitted by up to 4 peaks P$_1$ through P$_4$, each having a Voigt line shape which is the convolution of a Lorentzian with a Gaussian function. The temperature-dependent peak energies and areas are plotted in Figure 2b and Figure 2c respectively. At room temperature, P$_1$ has the highest peak energy. As the temperature is decreased, P$_1$ is no longer visible in the PL spectra. We leave Peak 1 for further discussion in Sec. 4 due to its limited data points in the temperature-dependent PL measurements. The energy of P$_2$ is 1.062eV at room temperature, close to the A exciton energy (1.07eV) in Ref (19). Consistent with previous studies,[20],[21] we identify the highest peak P$_2$ as the neutral exciton emission, originating from the direct transitions in the K or K' valley. The P$_2$ energies increase as temperature decreases. They are fitted by the Varshni Equation[22] $E_g(T) = E_g(0) - \frac{\alpha T^2}{T+\beta}$, which is typically used to describe the temperature dependence of semiconductor bandgaps. We obtain $E_g(0)$=1.13eV, $\alpha$=2.95×10$^{-4}$eV/K, and $\beta$=100K. We notice our parameters are slightly different from the ones obtained by Helmrich et al.[21] Such difference is believed to be a result of the different substrates used. In our case, the PDMS has a different thermal expansion coefficient from SiO$_2$. And the bilayer MoTe$_2$ could slide with respect to PDMS. Consequently, the peaks do not shift as much when the temperature is decreased. We conduct pump power-dependent PL measurements at 77K. The laser power at the sample is increased from 18.5µW up to 10mW. The corresponding PL spectra are shown in Figure 2d. Multiple Gaussian peaks are used to fit the spectra. Fitting details can be found in the Supporting Information (Figure S1). The pump power-dependent peak energies and areas are plotted in Figure 2e and 2f, respectively. As is shown in Figure 2e, the energy of $P_2$ decreases by 10meV as excitation density increases. Such redshift can be explained by a combined effect of bandgap renormalization and plasma screening.[23] The light-out-light-in (L-L) curve of the neutral exciton ($P_2$) in Figure 2f has the largest slope of 0.9 in the low pumping regime, close to 1, which is the slope for typical neutral exciton emissions. At pumping power higher than 0.74mW, the curves become sublinear, likely due to other non-linear processes such as bi-exciton formation or exciton-exciton annihilation, which is beyond the scope of this work.

Peak 3 ($P_3$) is centered at 1.028eV at room temperature, 35meV lower than $P_2$. It is not resolved as a separate peak at room temperature in previous studies.[20],[24] Its energies have a similar increasing trend as $P_2$ when temperature decreases, shown in Figure 2b. Its peak energy and intensity follow the same trend as $P_2$ when pumping increases, shown in Figure 2e and 2f. This suggests that the origin of $P_3$ is similar to that of $P_2$ or related to the direct transition in the K (K′) valley. We attribute $P_3$ to the trion emission, consistent with the trion binding energy of 24-27meV measured in monolayer $MoTe_2$.[25] The trion binding energy in bilayer $MoTe_2$ decreases from 35meV at RT to 20.5meV at 77K based on our measurement. As temperature decreases, trion emission is enhanced in Figure 2c, as a result of fewer trions ionized by the decreasing thermal energy.

Finally, we look at the lowest energy peak, Peak 4 ($P_4$). It centers around 0.96eV at room temperature and becomes more pronounced as the temperature decreases. It has a very broad peak with an FWHM of 214meV at 77K. And the slope of the L-L curve in Figure 2f is 0.813. With these observations, we tentatively assign it as the phonon-assisted indirect transition from $\Lambda_C$-$K_V$ corresponding to an indirect exciton. The indirect exciton emission increases with a decrease in temperatures. In general, PL due to indirect exciton recombination is determined by exciton and phonon population factors. Even though the phonon-assisted transitions comprise emission and absorption of phonons, the phonon emission dominates at low temperatures and hardly changes with the thermal populations. In that case, the temperature dependence of the PL is dominated by the exciton population factors. To this end, a detailed theoretical analysis is given in Ref (26), which also motivated our use of the fit formula, Eq. 1 for the area ratio of direct ($P_2$) to indirect ($P_4$) exciton emissions in Figure 4d. As shown by our theoretical calculations (details in the next section), the $\Lambda$-K bandgap is the smallest bandgap for bilayer $MoTe_2$. In addition, the $\Lambda$-K excitons are increasingly populated with decreasing temperature. These features agree with what is expected from the indirect $\Lambda_C$-$K_V$ transition.

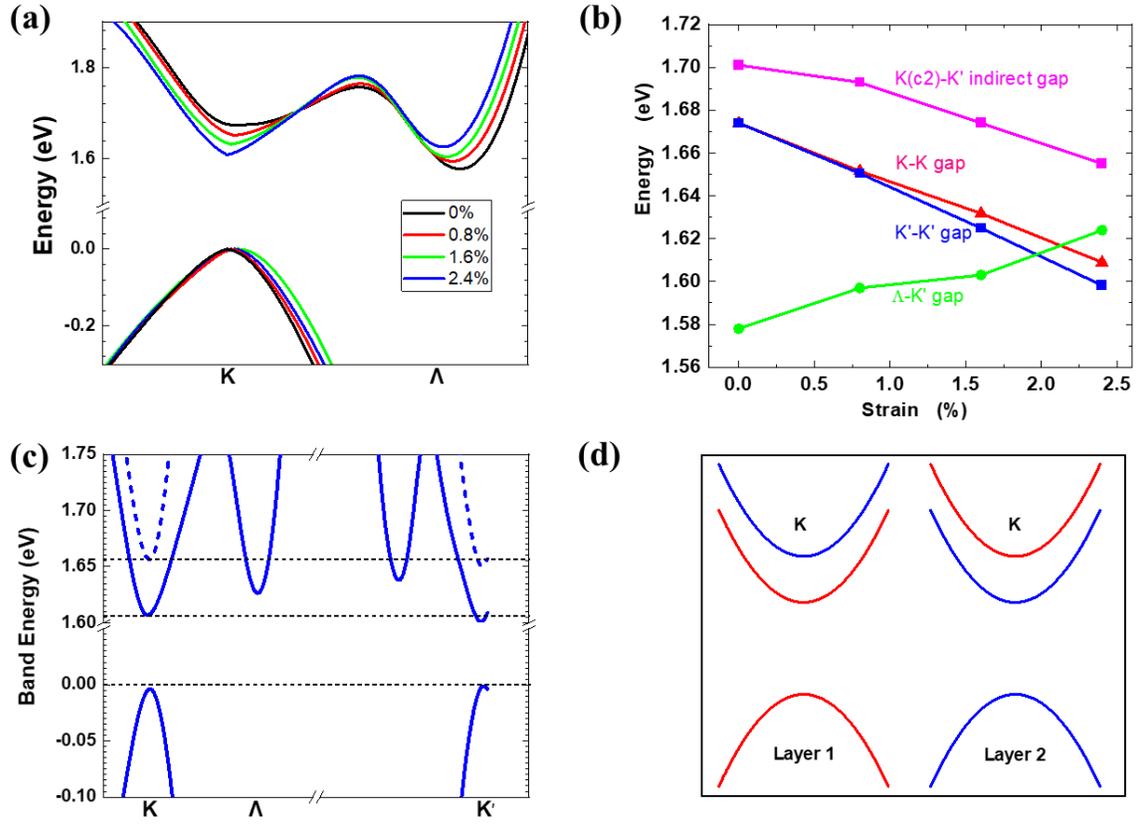

**Figure 3.** Theoretical Calculation of Band-Structures. (a) Electronic band structure for bilayer MoTe$_2$ at 0, 0.8%, 1.6%, 2,4% strain using DFT+GW calculation. Only the highest valence band and the lowest conduction band at K and Λ valleys are shown here for simplicity. (b) Bandgap energies for different transitions of bilayer MoTe$_2$. (c) Electronic band structure at selected k-points for bilayer MoTe$_2$ at 2.4% strain, dashed lines represent the second-lowest conduction band. (d) Schematic diagram of the band structures in the K valley. Energy states with the same spin are plotted in the same color.

Theoretical Results:

To more quantitatively understand the strain-induced changes in PL spectra, band structure calculations of the bilayer MoTe$_2$ were performed. Selected electronic bands of unstrained and strained bilayer MoTe$_2$ are compared in Figure 3a, while plots of the more comprehensive band structure are given in the Supporting Information (Figure S3). The valence band has its maximum value at K and K′ points with or without strain. K and K′ are degenerate in energy without strain. When no strain is applied, the conduction band has its minimum at the Λ point, indicating bilayer MoTe$_2$ is an indirect bandgap material, consistent with other calculations in literature.(27) As strain is applied, we see that the conduction band minimum at K and K′ decreases, accompanied by the increase of the minimum at Λ with the increase of strain levels. In other words, strain makes the bandgap of bilayer MoTe$_2$ more direct. Transition energies at selective k-points are extracted and plotted in Figure 3b. At 0% strain, the indirect transition from Λ in the conduction band to K′ in the valence band (green line) has the lowest energy of 1.578eV. Direct transition at K or K′ valley (red/blue lines) has a value of 1.674eV, leading to a Λ-K difference of ~ 96meV, among the smallest for bilayer TMDCs.[15]-[17] As strain is increased, Λ-K′ gap increases while both K-K and K′-K′ gaps decrease. Λ-K′ gap and K′-K′ gap cross over at strain values of approximately 2%, theoretically indicating a strain-induced indirect-direct bandgap transition in bilayer MoTe$_2$. The maximum strain (1.72%) in our experiment is quite close to the theoretical transition point.

Figure 3c shows the band structure in the larger range of wavevector at 2.4% strain with also the inclusion of the split-off conduction band due to spin-orbital coupling (SOC). We see a clear lifting of degeneracy between K and K′ due to the strain-induced symmetry breaking. Both of the conduction band edges are lower at K′ than at K, while the top of the valence band is slightly higher at K′ than at K. As a result, the bandgap is smaller (by 10meV) at K′. The degeneracy of K and K′ valleys has been lifted by the uniaxial strain.[28],[29] Such a lifting of the K-K′ degeneracy could lead to a preferential occupation of the K′ valley, leading potentially to steady-state valley polarization, especially at higher strain levels with larger K-K′ splitting.

Analysis Based on Experiment-Theory Comparison:

So far, Peak 2 has been attributed to the neutral exciton emission, Peak 3 to the trion emission, and Peak 4 to the indirect exciton emission, based on the discussion in the experiment result section. We also like to further discuss the origin of P$_1$ in Figure 2. In bilayer TMDCs with AB stacking, both spin and valley are coupled to the layer pseudospin.[30] In a given valley, spin configuration is locked to the layer index. We show the widely-adopted schematic spin polarized band structures in Figure 3d. States with the same spin are plotted in the same color. Contrary to WSe$_2$, in Mo-based TMDCs, the upper conduction band (CB2) in layer 2 has the same spin states with the highest valence band in layer 1 (see Figure 3d). With interlayer hopping, holes in layer 2 can form a bright exciton with electrons in CB2 in layer 1. Such spatially indirect exciton has been reported in other multilayer TMDCs like MoS$_2$,[31]-[33] MoSe$_2$[34], WSe$_2$[35] and heterobilayers.[36],[37] But so far they have not been observed in any room temperature PL spectra, likely due to the low quantum efficiency in those indirect gap multilayer TMDCs. In bilayer MoTe$_2$, the energy separation(ΔE1) between the two lowest conduction bands at K valley is 27meV at 0 strain, close to the separation (33.9meV) we see between P$_1$ and P$_2$ in Figure 2b. This suggests the association of P$_1$ with the interlayer exciton emission, whose energy should be higher than the intralayer K$_C$-K$_V$ exciton (P$_2$), as shown in Figure 3d. The near degeneracy of Λ and K in bilayer MoTe$_2$ results in more direct transitions in the K valley, compared to other TMDCs. This would also explain the fact that P$_1$ only exists in RT spectra. As temperature decreases, the interlayer indirect excitons are less populated. Theoretically, P$_1$ can also originate from Kc2-K′v transitions, like the momentum-dark excitons observed in monolayer[26] and bilayer[38] WSe$_2$. The intervalley transition requires emission or absorption of a phonon for conservation of energy and momentum. In this paper, we do not have enough evidence to completely rule out this possibility. But we consider that the interlayer exciton, which involves only intravalley recombination, is more likely the case here.

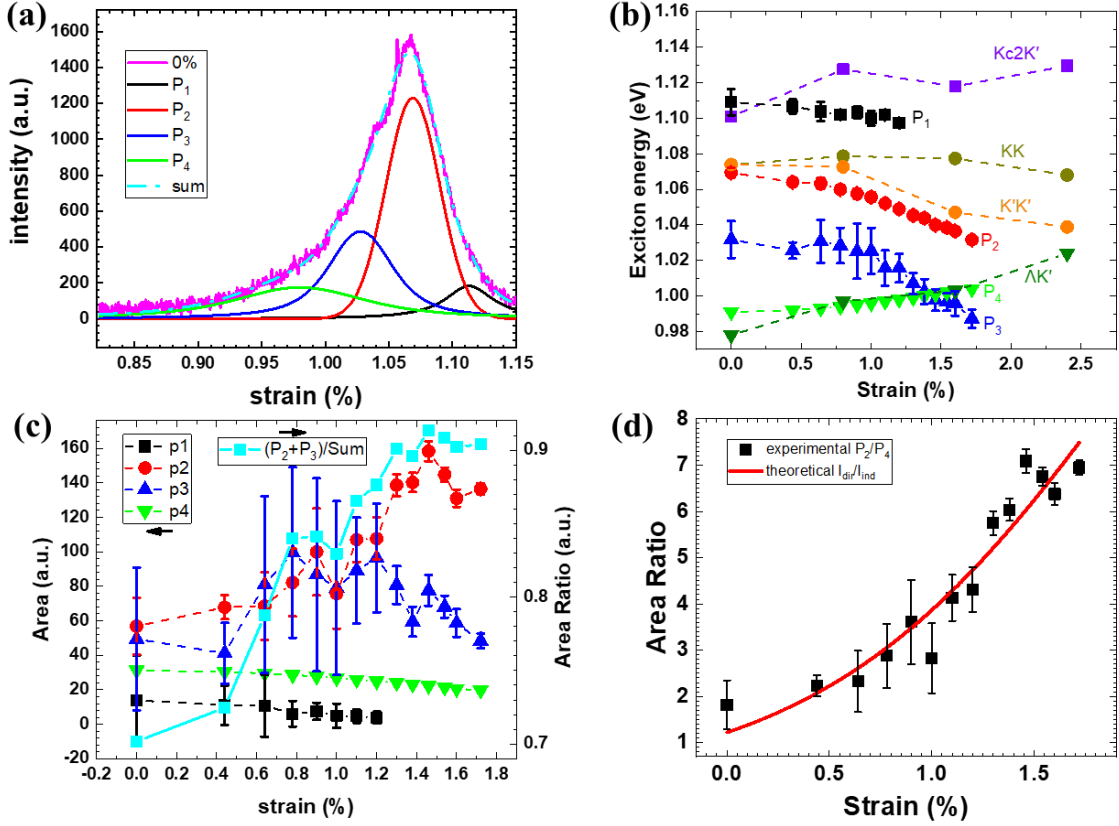

**Figure 4.** Strain-Dependent Spectral Features (a) PL spectra of bilayer MoTe$_2$ at room temperature fitted with multiple Voigt peaks at 0% strain. (b) Strain dependence of peak energies of Peak 1-4 and the theoretically estimated energies of different excitonic species. (c) Left y axis: Integrated intensity of each peak vs. strain; Right y axis: Ratio of the direct transition (P$_2$+P$_3$) over the total emission. (d) The experimental peak area ratio P$_2$/P$_4$ (black squares) and the fitting (red line) to the expression of $I_{dir}/I_{ind}$ in Eq. (1) to determine the coefficients a and b. The error bars in (b), (c), (d) represent one standard deviation in the least-squares fits.

Similar to the multi-peak Voigt fittings in Figure 2a, PL spectra at varying levels of strain at room temperature are fitted. An example of such fitting is shown in Figure 4a at 0% strain. Details of the fitting and corresponding peaks for the rest of the strain levels are shown in the Supporting Information S4.

The strain dependence of peak energies for the four peaks are shown in Figure 4b together with the theoretical energies of the K$_{c2}$K′, KK, K′K′, ΛK′ excitons (in this notation, the first state refers to the conduction band and the second to the valence band, e.g. ΛK′ refers to the transition from the lowest conduction band at Λ to the highest valence band at K′). In order to compute the exciton energies, we solve the Wannier equation for TMDCs within a 2-band model for hyperbolic bands with various mass parameters. For given strain, we then associate the band structure obtained from the first-principles calculation with the hyperbolic band that approximately reproduces the first-principles calculation band structure, and thus obtain an estimate for the exciton energy from the combination of the first-principles band structure and the solution of the Wannier equation. More details are given in Section 5 of the Supporting Information.

From Figure 4b we see that, without strain, $P_1$ corresponds to the theoretical $K_{C2}K'$ exciton with a difference of 8meV. We see that the peak energy of $P_1$ slowly decreases with strain while that of $K_{C2}K'$ fluctuates around 1.11eV. This deviation can result from the error in the estimation of the exciton binding energy or uncertainty in the fitting of $P_1$, given it is the weakest emission, especially at high strain levels. At zero strain, $P_2$ corresponds to KK or K'K' exciton (both are degenerate at zero strain) with a difference of only 5meV. $P_2$ and $P_3$ both redshift at the same rate (~ -28meV/%) as strain increases, as a result of the shrinking K-K and K'-K' gap. This further verifies that $P_2$ corresponds to the neutral-exciton emission within the K or K' valley, while $P_3$ corresponds to the associated trion emission, whose theoretical energy is not included in the calculation. After including the binding energy, we see a larger split between KK and K'K' exciton energies (30meV) at 2.4% strain compared to their bandgap difference (10meV). The trend of $P_2$ compares well with the theoretical value of the K'K' exciton within 0-1.6%. This indicates that at zero and small strain values, $P_2$ contains both KK and K'K' exciton. And as strain increases, the more populated K' valley makes the K'K' exciton emission dominate in $P_2$. We cannot rule out the possibility that the KK exciton emission is partially merged into $P_1$ at higher strain levels. Finally, $P_4$ corresponds to the ΛK' exciton with a difference of 13meV at 0 strain. The trend of the theoretical ΛK' exciton matches well with that of $P_4$. This again verifies the assignment of $P_4$ in the previous discussion about Figure 3b. We notice that the theoretical exciton energies for ΛK' and K'K' do not cross before the strain level of 2.4%, as opposed to their bandgap energies which cross before 2% strain (see Figure 3b). The experimental data show that the energy of $P_4$ is always lower than that of $P_2$. But it is clear from Figure 4b that both the theoretical and experimental results show the same consistent trend of indirect to direct transition. Our data shows that the maximum strain achieved in our experiment is very close to the theoretical indirect-direct transition point.

Figure 4c shows the integrated intensities of the four peaks as a function of the strain. We see that $P_2$ shows the most pronounced and almost monotonic increase with strain among all three peaks and is the main reason for the increase in the total PL intensity. This is consistent with the trend of the theoretical prediction of the indirect-direct band gap transition. As can be inferred from Figure 3a and 3c, the exciton population of the K' valley increases at the expense of the Λ valley population as strain increases, leading to more K'K' exciton population. Both direct exciton densities and recombination rate increase as strain modifies the band structure of the bilayer $MoTe_2$. Notably, the trion emission intensity reaches its maximum around the strain of 0.74%. This is reasonable as trions require the existence of excess electrons or holes which are limited as the bilayer $MoTe_2$ is sealed in PDMS. We also plot the ratio of direct transition ($P_2+P_3$) to the overall emission in Figure 4c. This ratio increases from 0.7 to over 0.9 as strain increases. The tensile strain has almost changed the bilayer $MoTe_2$ to direct gap material with 90% of emission coming from the direct transitions. This again verifies that the maximum strain value achieved here is very close to the required strain level for the indirect to direct transition. Above 1.5% of strain, both the exciton emission and total emission decrease with strain. Finally, we see that the intensity of $P_1$ gradually decreases with strain, and $P_1$ disappears at 1.2%. This can be explained by the increasing energy separation between $P_1$ and $P_2$, as predicted by the theoretical energy difference ($\Delta E1$) of the two conduction bands (see Figure 3b) at K as well. A detailed comparison of experimental peak area $P_1/P_2$ and its theoretical value can be found in the Supporting Information (Figure S7).

It is worth noting that the maximal strain value achievable in our experiment was limited by the mechanical set-up. We found that the spectral peaks stopped shifting for strain levels beyond 1.72%. It could be due to slippage of the $MoTe_2$ against the PDMS layers after a certain level of strain. As we pointed out, over 90% of the PL is already from the emission of the direct excitons under the current maximal strain applied. With better PDMS or other methods to improve the adhesiveness between $MoTe_2$ and the PDMS, we believe the overall transition to a direct bandgap could be further improved.

To further verify our peak assignment and to confirm the indirect-to-direct transition, we compare the PL intensity ratio of direct ($P_2$) to indirect ($P_4$) exciton emissions in Figure 4d. The experimental area ratio obtained from Figure 4c is compared with the fit formula

$$\frac{I_{dir}}{I_{ind}} = \frac{1}{a+b exp(\frac{\Delta E}{k_B T})} \qquad (1)$$

. Here $\Delta E$ is the energy difference between K′K′ and ΛK′ excitons as given in Figure 4b. This fit formula is motivated by the microscopic theory of indirect-exciton photoluminescence given as Eqs. (2) and (3) in Ref (26) , assuming that all sums in those equations contain only one state, and that the phonon frequency is much smaller than the energetic difference between direct and indirect excitons. The value for $\Delta E(eV)=-0.034\varepsilon+0.099$ is obtained by fitting the strain dependent energy difference of the two excitons (see Supporting Information Figure S8). The linear coefficient of -0.034 or -34meV/% is very close to the experimental value of -28meV/% obtained from a similar fitting. From Figure 4d, we see that the experimental area ratio of $P_2/P_4$ increases dramatically with strain, and this is well modelled by the fit formula of Eq. (1), using the parameter values a=0.055, b=0.016. The value of the parameter 'a' indicates the contribution of direct transition in the overall linewidth. In this case, the ratio of radiative recombination inside the light cone over the total decay (sum of radiative recombination inside the light cone plus phonon-assisted recombination) is 0.055 at room temperature.

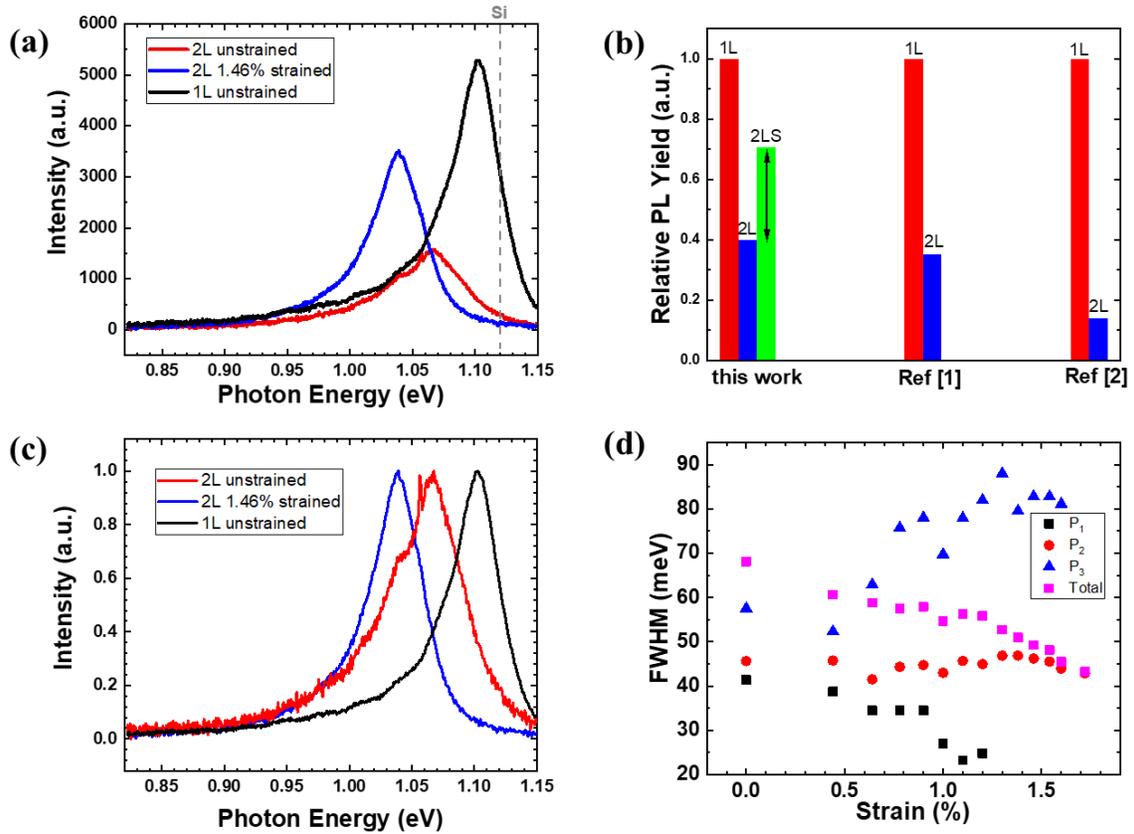

**Figure 5.** (a) PL spectra of the unstrained bilayer, strained bilayer, and unstrained monolayer MoTe$_2$. (b) Comparison of PL yield between monolayer and bilayer MoTe$_2$ from this work with those in the literature.[14],[19] (c) Normalized spectra in (a). (d) Resolved peak (with aforementioned peak assignment) linewidths (FWHM) and the full linewidth vs strain.

To have more quantitative measures of the strain effects on the PL qualities of bilayer MoTe$_2$, we compare the features of the PL spectrum of a strained bilayer MoTe$_2$ to those of typical unstrained monolayer and bilayer samples. As is shown in Figure 5a, the PL intensity of the strained bilayer has increased significantly and redshifted compared to the unstrained one. The PL intensity becomes more comparable to that of the unstrained monolayer. An additional benefit is that the redshifted emission of the strained bilayer MoTe$_2$ is further away from the silicon band edge (1.12eV at room temperature, marked by a grey dashed line). This makes the strained bilayer material even more attractive than both the unstrained bilayer and monolayer material for silicon-based devices due to the significantly reduced absorption. A more quantitative comparison of the relative PL yield can be found in Figure 5b. The PL yield (integrated intensity) of unstrained bilayer MoTe$_2$ in this work is 40% of that of the direct gap monolayer, comparable to the values extracted from Ref. (14) , (19) . However, the relative PL yield reaches 70% of the monolayer material or a 1.75× increase under strain. In Figure 5c, the normalized PL spectra for the three cases are compared. The strained bilayer MoTe$_2$ has a spectral linewidth of 43.2meV, much smaller than that of the unstrained bilayer (68.1meV). The reduced linewidth is even smaller than that of the monolayer (48.6meV). The strain-induced linewidth reduction is larger than that for other TMDCs[9]-[11] except for the bilayer WSe$_2$.[13] A detailed comparison can be found in the Supporting Information (see Figure S9). The linewidth contributions from various peaks are presented for the strained bilayer sample in Figure 5d. Although the linewidth of Peak 2 (the dominant exciton peak) is not significantly changed in the range of strain we studied, the overall linewidth is reduced by 36.6% as a result of changes of relative contributions from various peaks by comparing with Figure 4d. As the strain is increased, the exciton contribution (P$_2$) to PL emission increases rapidly (Figure 4d), while contributions from other peaks either decrease (P$_1$) or maintain the same (P$_3$). This leads to a dramatic reduction of the overall PL linewidth as seen in Figure 5c and 5d. As a result, the total linewidth of the PL spectrum eventually reflects the intrinsic linewidth of the neutral exciton emission at high levels of strain (compare for example 48.1meV for the total linewidth vs. 45.5 meV for the exciton linewidth at strain level of 1.54%).

**CONCLUSION**

In summary, we have systematically studied the effects of mechanical strain on the photoluminescence properties of bilayer MoTe$_2$, both experimentally and theoretically. Our results demonstrate a clear trend of indirect to direct bandgap within a reasonably low level of strain. The significant improvement of PL properties as reflected by the increased PL intensity and reduced linewidth demonstrates the unique effectiveness of strain engineering, especially on bilayer MoTe$_2$, due to a relatively small difference between the indirect and direct bandgaps in MoTe$_2$ compared to other bilayer TMDCs. Our detailed theory-experiment comparison also allowed us to explain the mechanism of PL intensity increase and the reduction of the PL linewidth. We found that strain plays the role of purifying the emission processes, making the intrinsic exciton emission the eventual dominant mechanism, leading to the enhanced emission and a reduced linewidth that is comparable to that of its monolayer counterpart. We believe that our method of study and the mechanisms identified could apply to other TMDC materials. From the application point of view, a strained bilayer is more preferable to both the unstrained bilayer and the monolayer counterparts, due to the significant redshift of the PL peak relative to the silicon absorption edge, making strained bilayer MoTe$_2$ one of the few 2D layered materials suitable for silicon-based photonic applications.

**METHODS**

**Experimental section:**

To prepare the samples, Scotch tape (3M 810-12BX-CA) was used to exfoliate MoTe$_2$ flakes from bulk material (2D Semiconductors). The tapes were then pressed against Polydimethylsiloxane (PDMS) (Dow Corning SYLGARD™ 184) films and slowly peeled off to transfer the flakes to PDMS. The stiffness of the PDMS was determined by the volume ratio between the base and the cure agent. Typically a 10:1 ratio is used. But for a higher efficiency of strain transfer, (see Supporting Information S9 for the discussion of the strain transfer efficiency) a more rigid PDMS is in favor.[39] Thus a 5:1 ratio was used instead. Bilayer MoTe$_2$ samples on PDMS were identified by their optical contrasts and subsequent PL measurements. Micro-PL measurements were conducted on a home-built system. A continuous-wave 633nm He-Ne laser was used as the excitation source. The incident laser was focused by a 50X NIR objective (Mitutoyo) with a spot size of around 2μm. PL signal was collected by an InGaAs detector (Symphony) coupled to the monochromator (Horiba iHR 320). The PDMS films were then cut into pieces of 1cm×2cm×0.5cm with bilayer $MoTe_2$ located at the center of each piece. The uniform geometry allows consistent strain control among different samples. The sample was then covered with another thin layer of PDMS deposited by drop-casting. The top layer of PDMS is around 10-20μm thick. It serves to clamp the 2D material, enabling larger strain to be transferred to them from the bent PDMS.

In the temperature-dependent PL measurement, a 10X objective was used to focus the pump laser and to collect the signal. Bilayer $MoTe_2$ samples on PDMS were put into a cryostat (Janis ST-500) cooled by flowing liquid nitrogen with a temperature controller (LakeShore model 331). The maximum actual laser power at the $MoTe_2$ plane was around 5mW. No sample damage was observed at such power level.

**Theoretical calculations:**

Band structure calculation of the bilayer $MoTe_2$ was performed by using the DFT+GW scheme as implemented in the VASP software package.[40] Given the ambient experimental condition in the present work, only the 2H phase (with hexagonal 2D lattice) for unstrained $MoTe_2$ was considered. The geometry optimization calculations used a plane-wave basis set with energy cutoff 400eV and optB86b functional,[41] which includes a non-local correction for van der Waals interaction. The lattice constant of the 2D hexagonal lattice used in the DFT+GW calculation was a=3.467 Å, which was obtained by the DFT level optimization with the settings listed above. A unit cell with c=28 Å was used in conjunction with a 10*10*1 k-point mesh to ensure the vacuum separation between periodical images. No symmetry constraints were applied in the calculations. The uniaxial strain was simulated by distortion of the unit cell in the ***a+b*** direction. We noted that qualitatively very similar results were obtained with distortion in the ***a-b*** direction.

Using the optimized geometries, the quasi-particle band structures of $MoTe_2$ were calculated at the $G_0W_0$ level. To this end, the initial wavefunctions incorporating 120 energy bands are calculated by using the hybrid functional HSE06[42] with SOC interaction included. After the $G_0W_0$ step, the band structures were interpolated by using the Wannier90 post-process package.[43] The GW calculations were tested by using 120 bands and 160 bands, respectively. At zero strain, the calculations by using both settings converged smoothly. The band structures obtained from both settings agree well with each other as well as with previous calculations.[14] For the strained lattices, we used a modified k-point path to sample the Brillouin zone with reduced symmetry and therefore reflect the lifted degeneracy (see Supporting Information Figure S2).

The exciton binding energy is obtained from the solution of the Wannier equation (similar to Bethe-Salpeter equation) with hyperbolic bands, where the parameters of the bands are taken from approximate fits of the DFT-GW band structure (see Supporting Information Figure S6).

**ACKNOWLEDGMENT**

We gratefully acknowledge the use of facilities at the Eyring Materials Center at Arizona State University, and support by the U.S. National Science Foundation (NSF) under grant No. ECCS-1807644 (ASU group) and DMR 1839570 (U of A group). The Paderborn group acknowledges funding from the Deutsche Forschungsgemeinschaft through project SCHU~1980/13 and the Heisenberg program (No.~270619725). Grants for computing time at the Paderborn Center for Parallel Computing and High Performance Computing (HPC) at the University of Arizona are also gratefully acknowledged.

**SUPPORTING INFORMATION**

Supporting Information Available: Fitting of the temperature- and pumping-dependent PL spectra at 77K; K-point selection in plotting the band structures under strained condition; Electronic band structures for bilayer $MoTe_2$ using DFT+GW calculation; Fitting of the strain dependent PL spectra at room temperature; Estimation of exciton binding energy; Theory-experiment comparison on the area ratio $P_1/P_2$; Estimated exciton energy difference $\Delta E$; Linewidth reduction comparison between different TMDCs under strain; Strain transfer efficiency between PDMS and $MoTe_2$.

# Supporting information:

## 1. Fitting of the temperature- and pumping-dependent PL spectra at 77K

Throughout this work, the multi-peak fitting was performed using program 'Fityk' (https://fityk.nieto.pl/). Initial parameters including center wavelengths, peak height, and lineshapes were set manually, followed by running the automatic Levenberg–Marquardt algorithm. The optimization process stops when the sum of the squares of the deviations is minimized. The peak functions and parameters were extracted and plotted for analysis. The error bars are one standard deviation for each parameter during the fitting, which are directly obtained from the program. For Figure 5d, the uncertainty of the linewidth of the strain dependent PL spectra is unavailable. In our fitting, we used multiple Voigt peaks, each of which is a convolution of a Lorentzian and a Gaussian. The uncertainty of the linewidth involves calculation of the covariance of the Lorentzian and Gaussian linewidths and their errors, which are not provided by the program (Fityk) we used.

In the process of our data analysis, we tried to fit the spectra using two, three, and four peaks. The parameters were determined through trials and iterations. It turned out that for the strain-dependent PL spectra at room temperature, only four-peak fitting gives reasonably accurate results and a consistent trend in all parameters (peak energies, linewidths, and intensities). In the two-peak fitting where only direct and indirect excitons were considered, the spectrum was either underfit at low strain values, or the parameters went through erratic changes at higher strain values. Based on our fitting experience of the strain-dependent spectra, 4 and 3 peaks were used for fitting the temperature- and pump-dependent spectra, respectively.

PL spectra at 77K are fitted with multiple Gaussian peaks shown in Figure S1. The peak positions are initially set following the values at 77K in Figure 2a. At low temperatures, peak 1 is absent. At pumping power below 740μW, peak 2 and peak 3 are used in the fitting. And at pumping power of 1.48mW and above, peak 4 is incorporated in the fitting.

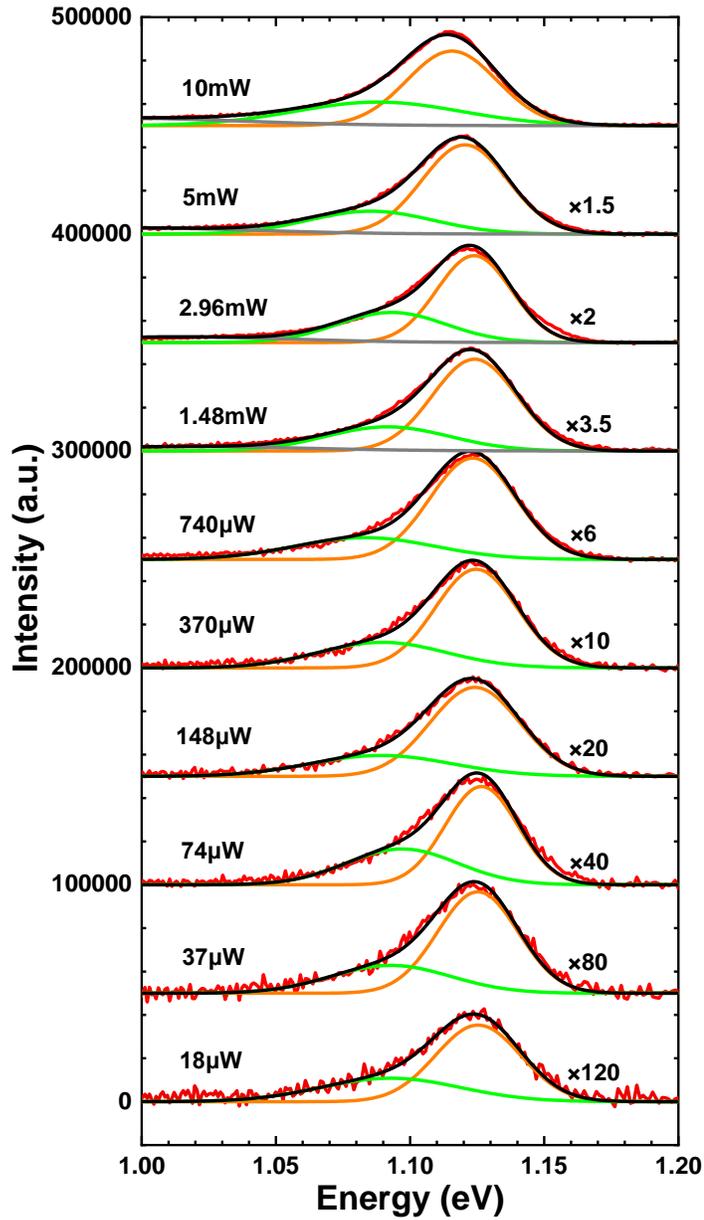

**Figure S1.** Pump power dependent PL spectra of bilayer MoTe$_2$ at 77K fitted with multiple Gaussian peaks. Orange line: peak 2; green line: peak 3; gray line: peak 4.

## 2. K-point selection in plotting the band structures under strained condition

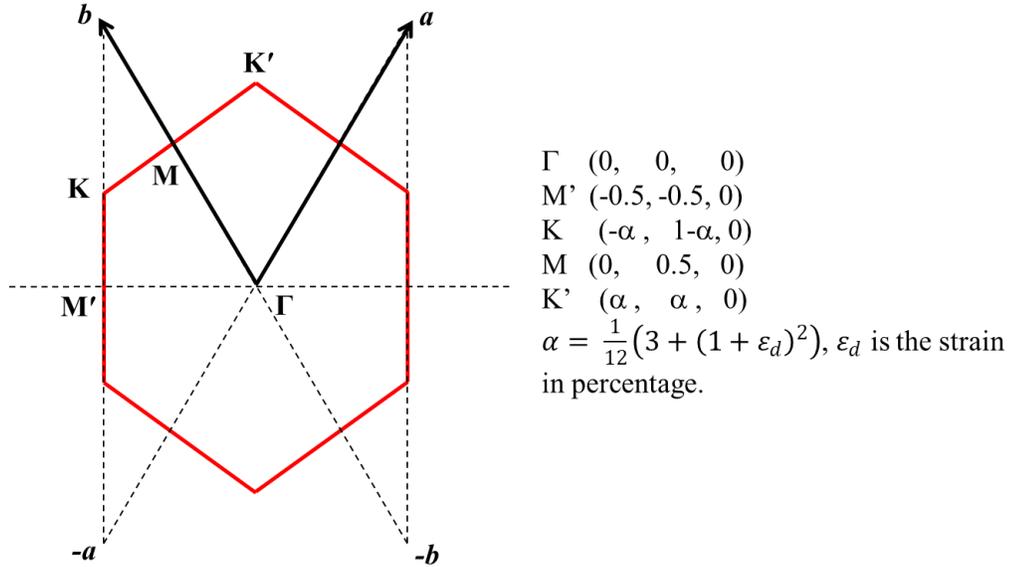

**Figure S2.** Illustration of the Brillouin zone of hexagonal lattice under strain.

Shown in Figure S2 is the red deformed hexagon where strain is applied on the ***a+b*** direction. The high-symmetry k-points, K and K', M and M', are no longer equivalent to those of the unstrained hexagonal Brillouin zone. ***a*** and ***b*** denote reciprocal lattice vectors. The correlation between strain ($\varepsilon_d$) along ***a+b*** direction and the special k-points used for the band structure calculation is presented. The k-path is selected as Γ-M′-K-Γ-M-K-Γ-K′. In the experiment the amount of lattice relaxation when strain is applied is unknown. In the z-direction in the calculations for the bilayer system, the system stays fully relaxed as the interlayer distance adjusts to each strain applied in the plane. For comparison with the results shown where the ***a-b*** direction stays fixed, we also let the lattice relax in the ***a-b*** direction. Qualitatively similar behavior is found for both scenarios but with relaxation, the indirect to direct bandgap transition of the bilayer system occurs for somewhat larger strain values.

## 3. Electronic band structures for bilayer MoTe$_2$ using DFT+GW calculation.

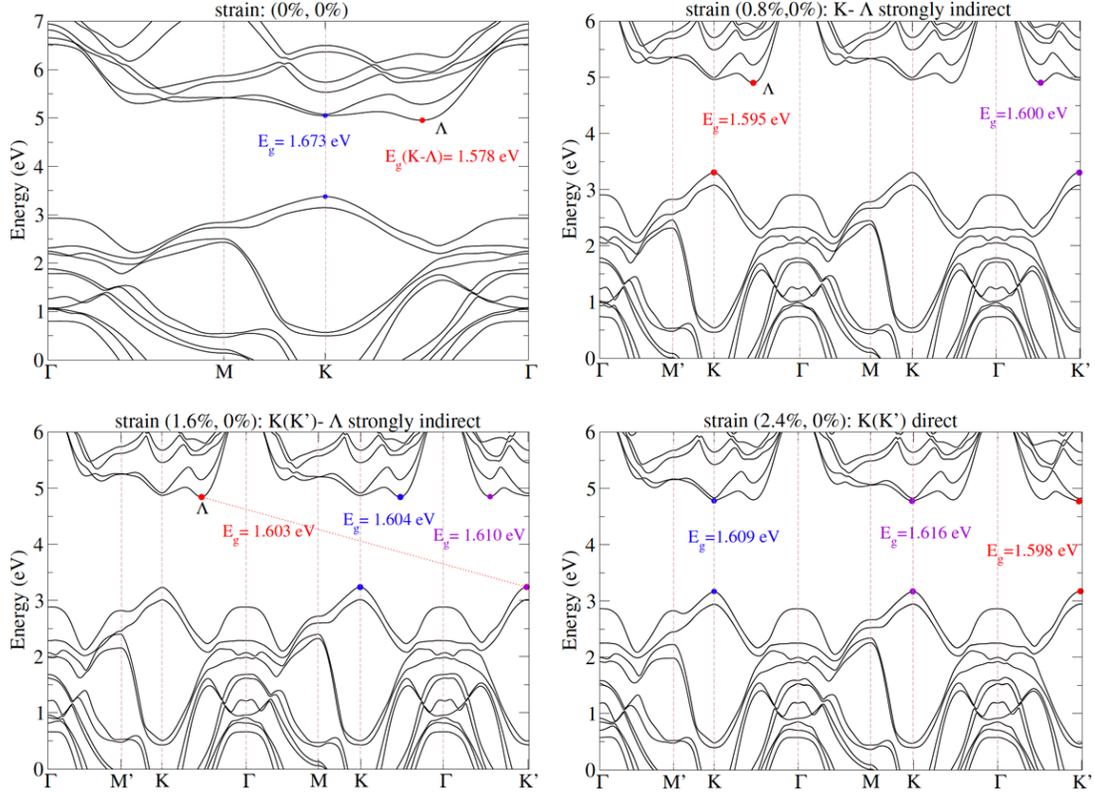

**Figure S3.** DFT+$G_0W_0$ calculated band structures of bi-layer $MoTe_2$ with different strain. The calculated band structures for different strain applied in ***a+b*** direction, as labelled on top of each panel (the first number in the bracket denotes the strain along ***a+b***, while the second number denotes strain along the orthogonal direction ***a-b***). The character of the fundamental band gap is also noted. For the bands of the structures with strain we show the k-point path **Γ**-M′-K-**Γ**-M-K-**Γ**-K′. In each panel the energy gap values are labelled corresponding to the k-points with same color.

Electronic band structures at four strain levels (0, 0.8%, 1.6% and 2.4%) are shown here in Figure S3 with the complete k-path included. Only the highest valence band and the two lowest conduction bands are shown.

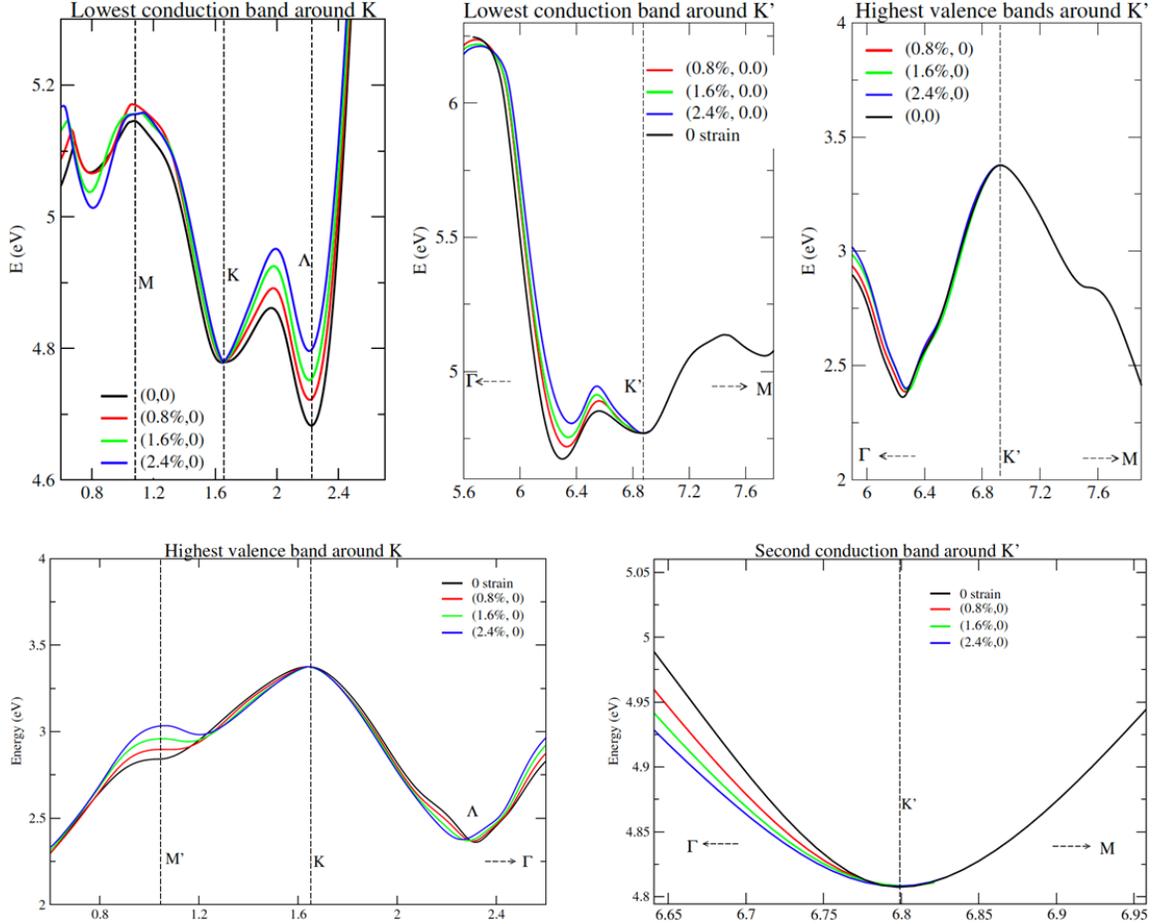

**Figure S4.** Local views of highest valence bands, lowest conduction bands, and second conduction bands around K or K′ points.

In Figure S4, the different bands are shifted in energy and in k-position such that minimum or maximum energy positions are aligned for easier comparison and estimation of the effective electron masses. The k-path directions are marked by the dash arrows as needed. In the upper middle and right figures, and in the lower right figure, the 0-strain local band structures around K′ is acquired by periodically extending the band structures around K, since K and K′ are equivalent for 0-strain.

## 4. Fitting of the strain dependent PL spectra at room temperature

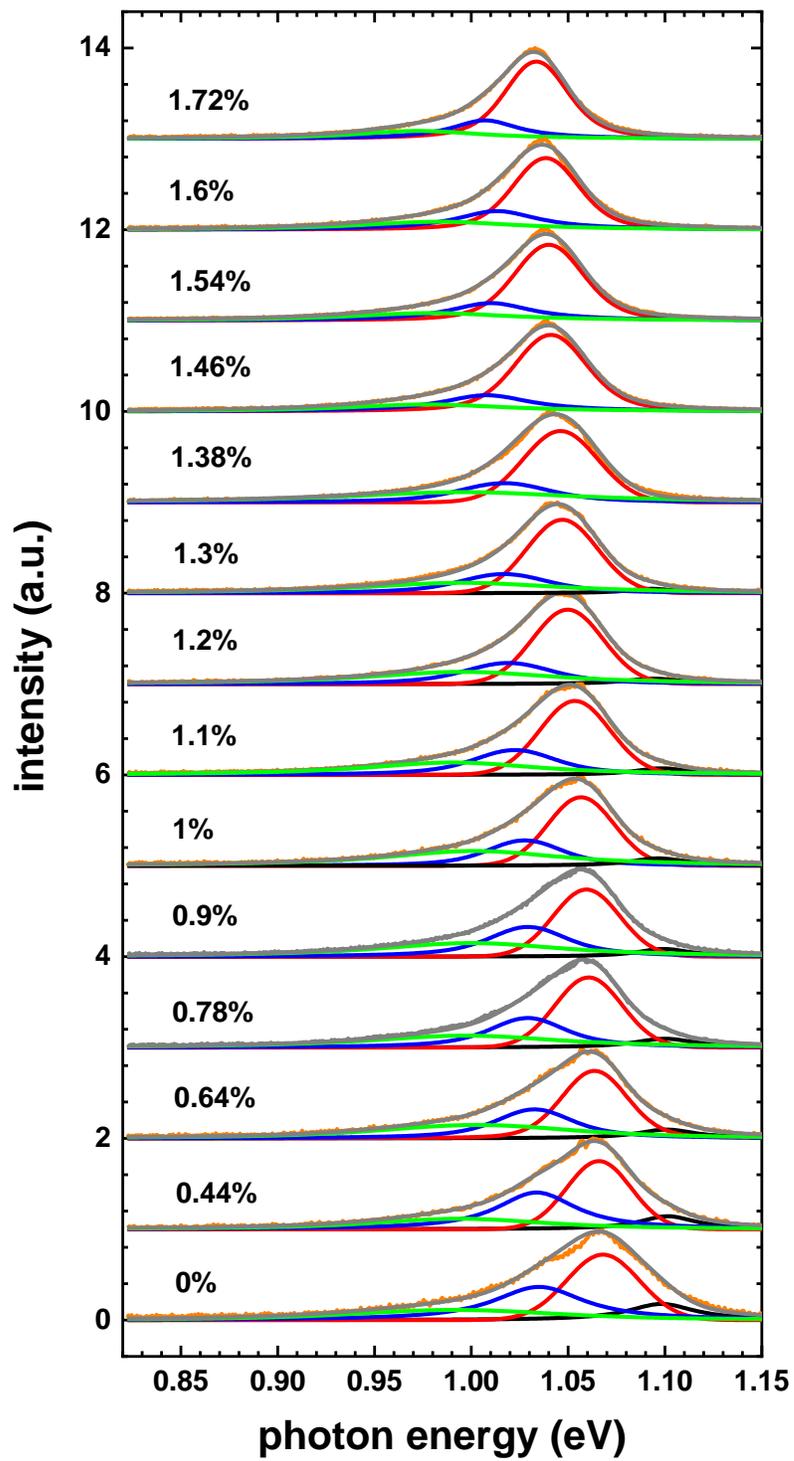

**Figure S5.** Normalized PL spectra of bilayer MoTe$_2$ fitted with multiple Voigt peaks at different strain levels. Black line: peak 1; red line: peak 2; blue line: peak 3; green line: peak 4; gray line: sum of P$_1$-P$_4$.

Strain dependent PL spectra at room temperature are fitted with multiple Voigt peaks shown in Figure S5. The peak positions are initially set following the values at 300K in Figure 2a. Four peaks (P$_1$-P$_4$) are used between 0-1.2%, and three peaks (P$_2$-P$_4$) are used for strain above 1.2%. Considering trion emission (P$_3$) takes up only a small portion in the overall emission at high strain values, it is possible that at those strain levels some of the direct exciton emission is mistakenly incorporated into the trion emission in the fitting, slightly broadening the trion peak.

Mathematically, the fitting can be done by the fitting program (Fityk) automatically with the number of initial parameters determined by the number of peaks. But practically, the automatic algorithm will typically give abrupt changes in extracted parameters, e.g. center wavelength, peak intensity, linewidths when strain value is changed. To avoid such unphysical situations and maintain the smooth changes of these fitting parameters with physical quantities such as strain, temperature or pumping, the peak position and intensity of one of the peaks (P$_4$) were chosen after several rounds of trials and errors. The reason P$_4$ was chosen is that it has the lowest spectral weight among all peaks, making it least likely to be simulated precisely by the program. Yet it plays an important role in the overall lineshape of the spectra, especially in the lower energy side. Additionally, the indirect-direct bandgap transition predicted by our calculation can give us an estimation of how this peak should evolve with respect to strain. This assumption is also consistent with the previous fitting of the unstrained PL spectra at room temperature. Now the validity of such semi-automatic peak fitting relies on the following criteria, (a) whether the other parameters (peak energy, intensity, linewidths) keep a continuous and consistent trend in the strain-increasing process; (b) whether the trend of the other parameters align with the theoretical prediction. It is shown in Figure 4(b), 4(d) and Figure S7 that all parameters match with the theoretical results well. The uncertainty in the fitting is generally kept at a low level. Although it is possible that better ways to fit this set of spectra might exist, the current fitting shown here represents our best understanding of the data.

## 5. Estimation of exciton binding energy

In this section we summarize the procedure used to obtain theoretical estimates for the exciton energies. As mentioned above, the band structure provides the band gap energies for each value of strain. It also provides the curvatures in the vicinity of the band extrema, which we parametrize by one parameter as follows. From the band structure plots, we read off the ratio of the band energy at a distance of $\Delta k = 0.25$ (in the units used in the band structure plots) away from the minimum over the energy at the minimum. In other words, each curvature is parametrized by

$$\frac{\varepsilon_{strained}(k_m + \Delta k)}{\varepsilon_{unstrained}(k_m + \Delta k)} \tag{1}$$

where $k_m$ denotes the wave vector at the extremum. For all cases where the curvature is different in the two directions shown, we take the average.

In order to associate each curvature with a corresponding exciton energy, we solve the Wannier equation for TMDs within a 2-band model for various mass parameters. In practice, we solve the equation for the wave vector dependent interband polarization $P(\mathbf{k})$ as a function of time, using a spectrally short pulse $\mathbf{E}(t)$ for the probe field, and then take the Fourier transform from time to frequency space and define the susceptibility as the ratio of the total polarization (i.e. $P(\mathbf{k}, \omega)$ summed over $\mathbf{k}$) divided by $\mathbf{E}(\omega)$. Specifically, we solve

$$i\hbar \frac{\partial}{\partial t} P(\mathbf{k}) = \left(\varepsilon_e(k) + \varepsilon_h(k) + E_g - i\gamma\right) P(\mathbf{k}) - \frac{1}{A} \sum_{\mathbf{k}} V^c_{\mathbf{k}-\mathbf{k}'} P(\mathbf{k}) - \mathbf{d}_{cv}(\mathbf{k}) \cdot \mathbf{E} \tag{2}$$

Here, the band energies of electrons, $\varepsilon_e(k)$, and holes, $\varepsilon_h(k)$, are taken to be equal, hyperbolic, and isotropic,

$$\varepsilon_a(k) = \frac{E_g}{2}\left(\sqrt{1 + 4\frac{a_L^2 t_h^2 k^2}{E_g^2}} - 1\right) \tag{3}$$

with $a = e, h$, $E_g$ denoting the bandgap, $a_L$ the lattice constant, and $t_h$ the hopping parameter, see, for example, Ref.(1), $\gamma$ is the dephasing and $\mathbf{d}_{cv}(\mathbf{k})$ the dipole matrix element (here taken to be $\mathbf{k}$-independent). The effective band mass is given by $m_e = \frac{\hbar^2 E_g}{2 a_L^2 t_h^2}$. We denote the unstrained mass by $m_e^{(0)}$. The numerical evaluations are performed for masses between $m_e^{(0)}$ and $0.5 m_e^{(0)}$. Our exciton model is isotropic, but the conduction band energy, and hence the effective mass, shown in the previous section is strongly anisotropic, with a large effective mass in the direction toward the $\Lambda$ valley. For this reason, we choose a relatively large value for $m_e^{(0)} = 1.6 m_0$, where $m_0$ is the electron mass in vacuum. Assuming equal masses for electrons and holes, the reduced mass is $m_r = m_e/2$.

The electron-hole Coulomb potential is

$$V^c(k) = \frac{2\pi e^2}{\epsilon_D(k) k} \tag{4}$$

where $e$ is the electron charge in vacuum, and we use the dielectric constant $\epsilon_D(k)$ taken from Ref.(2) wit $\ell_+ = \ell_- = 5d$ and $d = 0.6 \; nm$ in the notation of that reference.

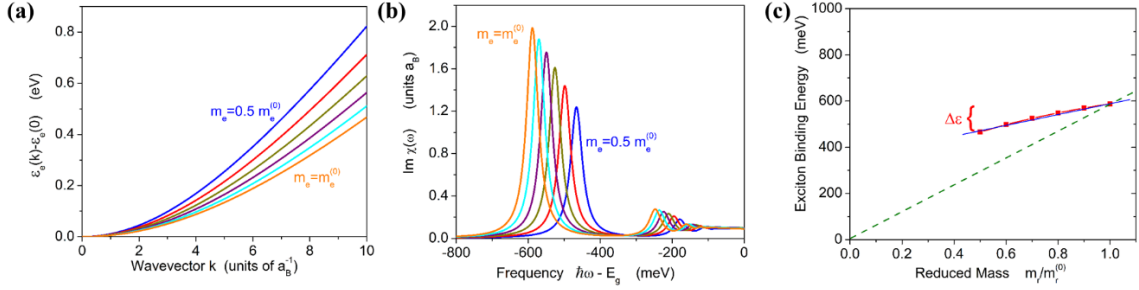

**Figure S6.** (a) Electronic band structure for 6 different mass parameters, used in the calculation of the exciton spectra. Here, $a_B$ denotes the exciton Bohr radius in three dimensions, its value being 2nm. (b) Exciton spectra given by the imaginary part of the linear optical susceptibility, for the 6 different mass parameters. (c) Summary of exciton binding energies for the 6 different mass parameters (red squares). The thin blue line is a linear approximation over the interval of mass parameters evaluated here. The green dashed line corresponds to an ideal hydrogen-like exciton model.

We show in Figure S6a the band energies entering the calculations. The corresponding spectra are shown in Figure S6b. We read off the binding energies from Figure S6b and summarize the results in Figure S6c. Over the range of masses evaluated here, the binding energy is approximately linearly related to the effective mass, changing by the amount of $-\Delta\varepsilon$ over this range,

$$E_B(m_r) = E_B\left(m_r^{(0)}\right) + \left(\frac{m_r}{m_r^{(0)}} - 1\right) 2\Delta\varepsilon \qquad (5)$$

We note that the linear relationship used here is different from what would be expected in a system with strictly hydrogen-like excitons, where the binding energy is proportional to the reduced mass (this is indicated by the dashed green line in Figure S6c.

Finally, the exciton energy is $E_g - E_B$ where the band gap energy for each interband transition considered in this paper follows from the band structure plot in the previous section.

## 6. Theory-experiment comparison on the area ratio $P_1/P_2$

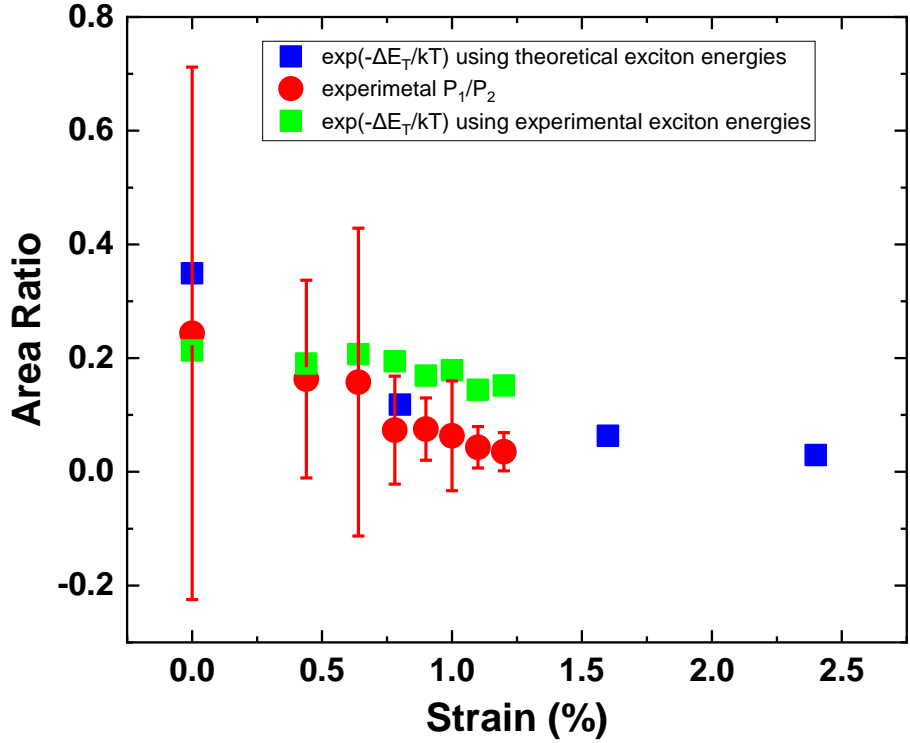

**Figure S7.** Comparison of the experimental peak area ratio $P_1/P_2$ and its theoretical value estimated by $\exp(-\Delta E_T/kT)$, where $\Delta E_T$ is the energy difference between the two lowest conduction bands at the K′ point. Error bars represent one standard deviation in the fitting, see Figure S5.

By considering the effects of carrier population only, the ratio of the peak intensities $P_1/P_2$ can be expressed by $\exp(-\Delta E_T/kT)$ using the Boltzmann distribution where $E_T$ is the energy difference between $K_{C2}K'$ exciton and $K'K'$ exciton. It is compared with the experimental value as shown in Figure S7. Here $E_T$ is obtained from both the theoretical and experimental values in Figure 4b. The experimental $P_1/P_2$ and the two calculated exponential values all show the same decreasing trend as strain increases. As the conduction band is further split off by strain, fewer carriers can populate the upper conduction band, leading to the decrease in the relative intensity of $P_1$. This is additional evidence suggesting that $P_1$ originates from CB2, indicating the possibility of strain-induced modification of the interlayer exciton emission.

## 7. Estimated exciton energy difference ΔE

The theoretical exciton energy difference ΔE is plotted in Figure S8 based on the values in Figure 4b. It shows a decreasing trend as strain increases. The black dots are fitted by the linear curve $\Delta E(eV)=-0.034\varepsilon+0.099$ represented by the red dashed line. The energy difference can be described by this relation well within 0-2.4% strain range.

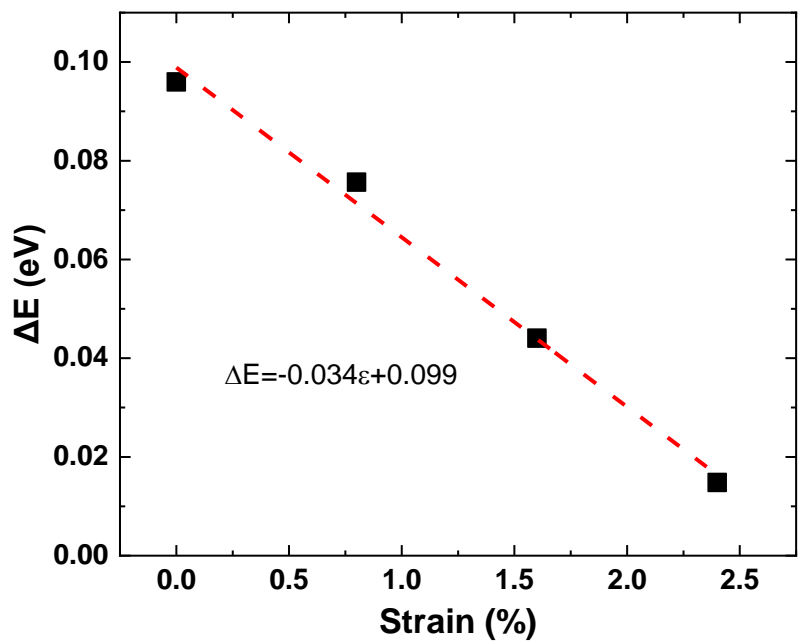

**Figure S8.** Theoretical energy difference between the K′K′ (direct) exciton and ΛK′ (indirect) exciton vs strain (black dots) and the linear fitting (red dashed line).

## 8. Linewidth reduction comparison between different TMDCs under strain

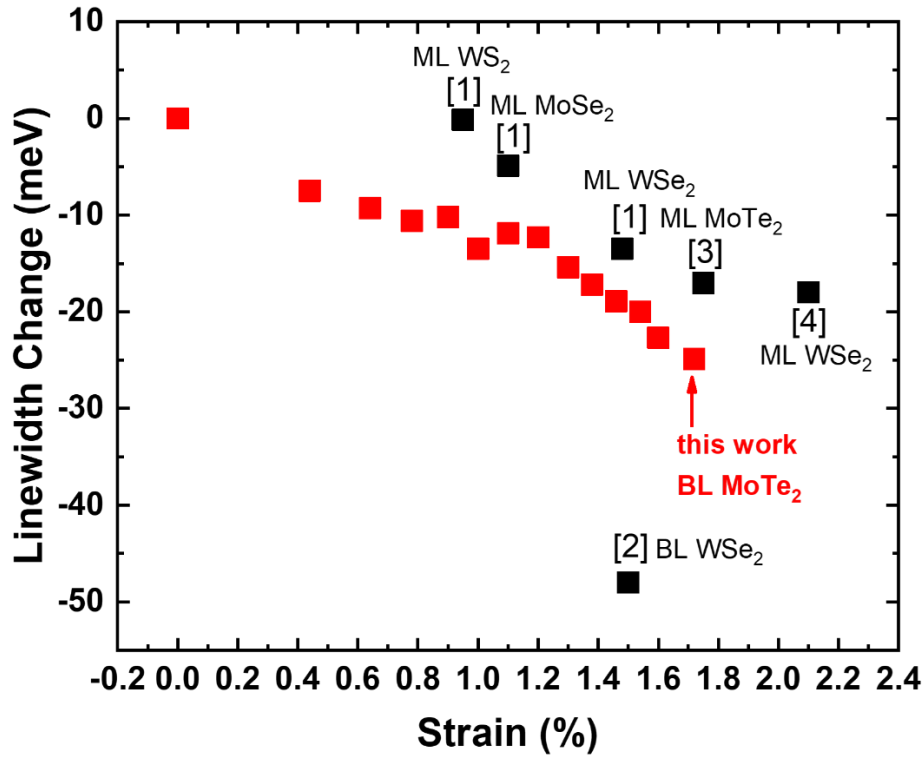

**Figure S9.** Comparison of linewidth reduction between different TMDCs, black dots: extracted from literature [1](1) ([3) [2]([4) [3]([5) [4]([6) , red dot: this work.

In this plot we have included all the strain-induced linewidth reduction observed in PL spectra in different TMDCs to the best of our knowledge. We find that except the complete indirect to direct band gap transition observed by Desai et al. in bilayer $WSe_2$,(4) our work has seen the second largest linewidth reduction (24.9meV) in 2D TMDCs, although the maximum strain value recorded in our case is at a median level compared to others' work. The large linewidth reduction in bilayer $MoTe_2$ and bilayer $WSe_2$ both originates from the indirect to direct band gap transition. The near degeneracy of ΛK and KK gap of bilayer $MoTe_2$ at 0 strain has limited its potential amount of linewidth reduction compared to $WSe_2$.

## 9. Strain transfer efficiency between PDMS and $MoTe_2$

In strain-related study of 2D materials, the strain value is typically estimated by calculating the structural deformation of the substrate. The uncertainty of the determined strain level experienced by the 2D material is the strain transfer efficiency. The loss of strain transfer between interfaces is well known and is a general issue in almost all strain-related work.

Stiffness[7] or Young's modulus[8] is known to be able to affect the strain-transfer efficiency. From our experience, we found that the base/curing agent ratio, which essentially changes the stiffness of the PDMS, affects strain-transfer efficiency sensitively. Our study suggests that the higher curing agent ratio produced PDMS with higher stiffness. And the ratio of 5:1 that we eventually used transfers strain much more effectively than the ratio of 10:1, the most commonly used one. We were also aware that sliding and debonding could occur during strain application. So we tried to minimize such issue by drop-casting another layer of PDMS. When damaging and de-bonding occurs, the local strain in the material will be released, which sets the strain back (or close) to zero instantly. And this has actually happened on some of our samples, on which we could see an abrupt shift of PL peak during the measurement. On the samples with which we showed the PL spectra (Figure 1 b,c), the peak shift was continuous and reversible. That means the strain transfer was efficient, and no damage occurred.

It is a legitimate concern that the true strain in the $MoTe_2$ layer might be smaller than that in the PDMS layer. However, since the observed changes of the PL spectra are substantial, the strain that causes those spectral modifications has to be substantial. Our theory (as well as most of the literature) predicts substantial PL changes to occur for strains on the order of 0.1 to 1%. Experimentally, both the maximum strain (1.72%) and the resulting rate of peak shifting (-28meV/%) in our case are comparable to all the values in the similar strain-related work. And the experimental value of -28meV/% is very close to the theoretical value of -34meV/%. They all give us high confidence of the validity of the data. We can be certain that our estimate is in the right order of magnitude. In other words, the true strain in the $MoTe_2$ layer might be a little bit smaller than the values given on our curves, but not much smaller.